\documentclass[12pt,preprint]{aastex}

\usepackage{graphics}

\newcommand{\be}{\begin{equation}}
\newcommand{\ee}{\end{equation}}

\shorttitle{Origin of radio emission from LLAGNs} \shortauthors{Wu
\&  Cao }

\begin{document}

\title{ Origin of radio emission from nearby low-luminosity active galactic nuclei}

\author{$\rm Qingwen\   Wu^{1,\ 2} \ and  \    Xinwu\  Cao^{1}$}
\affil{1. Shanghai Astronomical Observatory, Chinese Academy of
Sciences,
 Shanghai, 200030, China \\
2. Graduate School of Chinese Academy of Sciences, Beijing,
100039, China \\
Email: qwwu@shao.ac.cn\\
        cxw@shao.ac.cn}

  \clearpage
  \begin{abstract}
  We use the observational data in radio, optical and X-ray
  wavebands, for a sample of active galactic nuclei (AGNs) with measured
  black hole masses, to explore the origin of radio emission from
  nearby low-luminosity active galactic nuclei (LLAGNs). The
  maximal luminosity of an advection dominated accretion
  flow (ADAF) can be calculated for a given black hole mass, as
  there is a critical accretion rate $\dot{m}_{\rm crit}$,
  above which the ADAF is no longer present. We find that the radio luminosities are higher
   than the maximal luminosities expected from the
  ADAF model, for most sources in this sample. This implies that the radio
  emission is dominantly from the jets in these sources. The
  radio emission from a small fraction of the sources (15/60, hereafter referred to as radio-weak
  sources) in this sample can be explained by the ADAF model. However, comparing the observed multi-band
  emission data with the spectra calculated for the ADAF or ADIOS (adiabatic inflow-outflow
  solution) cases, we find that neither ADAF nor ADIOS models can
  reproduce the observed multi-band emission simultaneously, with reasonable magnetic field
  strengths,
   for these radio-weak sources. A
  variety of other possibilities are discussed, and we suggest
  that the radio emission is probably dominated by jet emission even
  in these radio-weak LLAGNs.

    \end{abstract}

\section{Introduction}
   There is increasing evidence for supermassive black holes in
    both distant luminous AGNs and nearby
    galaxies. Black hole accretion is thought to power AGNs, and the
    UV/optical bump observed in luminous quasars is naturally interpreted as blackbody
    emission from standard thin disks \citep{sm89}. A small
    fraction of quasars ($\sim$ 10 percent) are radio-loud, which are
    defined by radio-loudness $R\equiv L_{\nu}(\rm 6\ cm)$/$L_{\nu}(\rm
    4400\ \AA)>10$ \citep{ke89,vi92,st92,ke94,hi95}. High resolution radio
    observations show that the radio emission of these luminous radio-loud quasars is
    usually related to their relativistic jets.
     It is found that some nearby active
    galaxies share similar characteristics with those luminous
    quasars, but with relatively weaker broad-line emission and
    radio cores \citep{jb95,ho97,bbr00,f00,n01}. These low-luminosity
    active galactic nuclei (LLAGNs) are very common in the local
    universe. According
   to the Palomar survey \citep{hfs97}, more than 40 percent of all nearby
   galaxies are brighter than $B_{T}=12.5\ \rm mag$ and emit AGN-like
   optical spectra.

   Most of these LLAGNs are highly sub-Eddington
   systems, i.e., $L_{\rm bol}/L_{\rm Edd}<10^{-2}$ \citep{ho03}.
   Accreting at less than a few percent of the Eddington rate,
   the black hole may accrete via an ADAF \citep{ny94,ab95,ny95a,ny95b,ch95}.
   The ADAF model can successfully  explain most observational features of
    Sagittarius $\rm A^{*}$ at the center of our
    Galaxy \citep{nym95,mmk97,ma98,na98} and other LLAGNs, such as
    NGC4258 \citep{la96,gnb99}.

   \citet{la96} suggested that the ADAFs
   in most LLAGNs should be truncated to standard thin disks at certain radii, in order to explain the observed spectra.
   Several investigations were carried out to fit the spectra of
   LLAGNs, such as M81 and NGC4579,
   using the ADAF+SD (standard disk) models \citep{qd99}.
   This model can successfully explain most important features of LLAGNs, but
   underpredict the detected low frequency radio flux by more than one order of
   magnitude\citep{f04,ymf02,fm00}. They proposed that outflows/jets might be in these
   sources, and that the presence of such outflows/jets is supported by
   theoretical calculations on ADAFs as they have positive bernoulli constants \citep{ny94,ny95a}.
    Recently, the modified ADAF model, the powerful
    windy ADAF (adiabatic inflow-outflow solutions [ADIOSs])
    \citep{bb99},
    and
    convection dominated flows (CDAFs) \citep{ni00, spb99}, were proposed
     to explain the observational features of LLAGNs.
    Some LLAGNs have been observed and detected with the Very Long
    Baseline Array (VLBA) at different frequencies \citep{uh01,nwf01,au04},
    which implies that the jet should dominate the
    radio emission from the nuclei in some sources.
    The VLBI observations show that the radio core sizes are very small in most of these sources.

    In this paper, we use a sample of LLAGNs, with measured black hole
    masses, to explore the origin of their radio emission.

\section{The sample}
        The LLAGN sample used for present investigation is selected from \citet{ho02}.
    The sources in this sample, which have the most
    up-to-date estimated black hole masses, high angular resolution optical and
    radio measurements. This sample covers a wide range of nuclear activity, from
    nearby inactive nuclei to classical Seyfert 1 nuclei and luminous
    quasars. In our present investigation, we focus on the LLAGNs, and the Palomar-Green quasars
    are not included, as they are well-known
    luminous quasars. The optical emission from the host galaxies
    dominates over that from the nuclei for most LLAGNs. In some extreme cases, the nucleus
   accounts for merely 0.01 percent of the integrated light \citep{hp01}. The
    emission from the nuclei reflect the nature of the central
    engines in these LLAGNs.
     The core radio emission has been measured by the VLA at 6 cm, with beam size $\lesssim
    5^{''}$. The nuclear spectral power $P_{\rm 6, nuc}$ at 6 cm,
    computed from the observed flux density, represents the emission associated with the
    ``active'' (nonstellar) nucleus. Since a number of objects have type 2
    nuclei, which may be hidden from direct view, or at least appreciably extincted,
    the known relation between $\rm H\beta$ luminosity and
    $B$-band
    absolute magnitude for type 1 AGNs is used to estimate the
    $M^{'}_{B, \rm nuc}$ of these LLAGNs \citep{hp01}. \citet{ho02} introduced
   a modified radio-loudness $R^{'}\equiv L_{\nu, \rm c}^{'}(\rm 6\
   cm)$/$L_{\nu, \rm c}^{'}(\rm 4400\ \AA)$ to measure the radio
   activities in these LLAGNs, where $L_{\nu, \rm c}^{'}(\rm 6\
   cm)$ and $L_{\nu, \rm c}^{'}(\rm 4400\ \AA)$ are the core luminosities, as measured by
    the Very Large Array (VLA) and Hubble Space Telescope
   (HST),
    respectively. It is found that all these LLAGNs are radio-loud ($R^{'}>10$) by this definition, instead of
   radio-quiet in the conventional definition.

   X-ray observations can also provide a direct probe of the
   central
    engines, if the X-ray emission is not obscured. We searched
    the literature, and compiled the X-ray emission data measured at
    different energy bands, which were
    converted to $L_{\lambda}(2\ {\rm keV})$. For the sources
    without measured photon indices, we adopt the conventional
    assumption that photon index $\Gamma=2$. As a result, we have a sample of 62
    LLAGNs, of which 57 have measured optical nuclei, 60 have
    measured radio nuclei, and 57 have measured X-ray emission. All
    the data in this sample are listed in Table 1, and more details
    may be found in \citet{ho02}.

\section{Spectra of ADAF/ADIOS}
     In this work, we use the following dimensionless variables:
   \be
   m=\frac{M_{\rm bh}}{M_{\odot}},\  r=\frac{R}{R_{\rm Schw}},\  \dot{m}=\frac{\dot{M}}{\dot{M}_{\rm
   Edd}},
   \ee
   where the Schwarzschild radius is defined as
    \be
   R_{\rm Schw}=\frac{2GM_{\rm bh}}{c^{2}}=2.95\times10^{5}m \ \rm cm,
   \ee
   and the Eddington accretion rate is defined as
    \be
  \dot{M}_{\rm Edd}=\frac{L_{\rm Edd}}{\eta_{\rm eff}c^{2}}=1.39\times10^{18}m \rm\ g
  \  s^{-1}.
  \ee
  An accretion efficiency of  $\eta_{\rm eff}=0.1$ is adopted in this work \citep{fk92}.
    When the mass accretion rate is below a critical rate, $\dot{m}\lesssim 0.28\alpha^{2} $,
     where $\alpha$ is the viscosity parameter, the
    accretion flow
  is thought to be an optically thin
  ``advection-dominated'' accretion flow (ADAF) \citep{ny95b,m97}.
  We can calculate the ADAF spectrum, using the approach proposed
  in M97, if the  black hole mass $m$, the dimensionless accretion rate $\dot{m}$,
  the viscosity $\alpha$, and the ratio $\beta$ of the gas to total pressures,
  are specified. For the fixed black hole mass $m$, the ADAF
luminosities in different bands (i.e., the optical, radio and
X-ray) increase
 with $\alpha$.
    \citet{c02} showed that the maximal
 optical luminosity $L_{\lambda}^{\rm max}(4400 {\rm\ \AA})$ of an ADAF always requires
 $\beta=0.5$, i.e., equipartition between the magnetic and gas pressures.
  Our numerical results show that the maximal radio luminosity
 $L_{\lambda}^{\rm max}(5\ {\rm GHz})$ requires $\beta=0.5$, and that the maximal
 X-ray luminosity
 $L_{\lambda}^{\rm max}(2\ {\rm keV})$ requires $\beta\simeq0.7$, if all other parameters are
 fixed. By varying the dimensionless accretion rate $\dot{m}$,
 we can obtain the numerically
 maximal luminosity at a given wavelength $\lambda_{0}$, as a function of black hole mass $m$.

 For an ADIOS flow, the accretion rate is a function
  of  radius $r$, instead of being a constant independent of $r$ as in a
  pure ADAF model \citep{bb99,qn99}. A dimensionless
  accretion rate,  $\dot{m}=\dot{m}_{\rm out}(r/r_{\rm max})^{p}$, is assumed for the ADIOS flow, where
  $\dot{m}_{\rm out}$ is the mass accretion rate at $r_{\rm max}$, and $p$ is
  a parameter describing the strength of the winds \citep{bb99}. For strong
  winds (corresponding to a large $p$), the spectrum emitted from the flow is modified,
  as its structure is significantly  altered by the winds.
   Both synchrotron and bremsstrahlung emission
  decrease with increasing $p$, as a large $p$ leads to lower density and
  lower electron temperature in the inner region of the flow. An ADIOS is described by a
similar
 set of equations as for an ADAF, with an $r$-dependent accretion rate
  $\dot{m}(r)$  assumed. The spectral calculations for
  the ADIOS flows are presented in \citet{ccy02}, which are briefly
  summarized here (several errors in their paper have been fixed).

The  total heating rate $Q^{+}$ of the ADIOS flow is given by
 \be
Q^{+}=9.39\times10^{38}\frac{1-\beta}{f}c_{3}m\dot{m}_{\rm
out}r^{-p}_{\rm max}\frac{1}{p-1}(r^{p-1}_{\rm max}-r^{p-1}_{\rm
min}), \ee
 where $p<1$, $r_{\rm max}$ is the outer radius of the ADIOS
 flow, which extends to $r_{\rm min}$.

The total ion-electron heating rate for the electrons in the flow
is
 \be
Q^{\rm ie}=1.2\times10^{38}g(\theta_{\rm
e})\alpha^{-2}c^{-2}_{1}c_{3}\beta m\dot{m}^{2}_{\rm
out}r^{-2p}_{\rm max}\frac{1}{2p-1}\times(r^{2p-1}_{\rm
max}-r^{2p-1}_{\rm min}), \ee
 where $p\neq1/2$, $c_{1}$, $c_{3}$  are the constants as defined in
 \citet{ny95b}, and $g(\theta_{\rm e})$ is given by Eq. (11) in M97.

 As suggested by \citet{ny95b}, it is a good approximation to assume $\chi_{M}$ and $T_{\rm e}$
 to be constant in the flow for spectral calculations.
 For a given $\chi_{M}$, the cut-off frequency of cyclosynchrotron emission at each radius is given by
\be \nu_{\rm c}=s_{1}s_{2}m^{-1/2}\dot{m}_{\rm
out}^{1/2}r^{-p/2}_{\rm max}T_{\rm e}^{2}r^{(2p-5)/4}, \ee where
$s_{1}=1.42\times10^{9}\alpha^{-1/2}(1-\beta)^{1/2}c_{1}^{-1/2}c_{3}^{1/2}$,
$s_{2}=1.19\times10^{-13}\chi_{M}$, $\chi_{M}\equiv2\nu/3\nu_{\rm
b}\theta_{e}^{2}$, $\nu_{\rm b}\equiv eB/2\pi m_{\rm e}c$, and
$e$, $B$, $m_{\rm e}$ are electron charge, the magnetic field
strength and electron mass respectively.
 The peak frequency of cyclosynchrotron emission is originated at $r_{\rm
min}$, \be \nu_{\rm p}=s_{1}s_{2}m^{-1/2}\dot{m}_{\rm
out}^{1/2}r^{-p/2}_{\rm max}T_{\rm e}^{2}r_{\rm min}^{(2p-5)/4}.
\ee The synchrotron spectrum of the ADIOS flow is then described
by
 \be
 L_{\nu}^{\rm sync}=s_{3}(s_{1}s_{2})^{8/(5-2p)} m^{(4p-6)/(2p-5)}
 \dot{m}_{\rm out}^{4/(5-2p)}r_{\rm max}^{4p/(2p-5)}
 T_{\rm e}^{(21-2p)/(5-2p)} \nu^{(4p-2)/(2p-5)},
\ee where $s_{3}=1.05\times10^{-24}$.
   We can obtain the radio luminosity at peak frequency $\nu_{\rm p}$,
  \be
\nu_{\rm p}L_{\nu_{\rm p}}^{\rm
sync}=s_{3}(s_{1}s_{2})^{3}m^{1/2}\dot{m}_{\rm out}^{3/2} r_{\rm
max}^{-3p/2}T_{\rm e}^{7}r_{\rm min}^{(6p-7)/4}. \ee The total
synchrotron power is approximately given by
 \be P_{\rm sync}\simeq
\int_{0}^{\nu_{\rm p}}L_{\nu}^{\rm sync}d\nu =
\frac{2p-5}{6p-7}\nu_{\rm p}L_{\nu_{\rm p}}^{\rm sync}. \ee
The
total power of the bremsstrahlung emission and its spectrum are
given by
\be P_{\rm
brem}=4.74\times10^{34}\alpha^{-2}c_{1}^{-2}m\dot{m}^{2}_{\rm out}
r^{-2p}_{\rm max}F(\theta_{\rm e})\times\left[\frac{1}{2p}(r_{\rm
max}^{2p}-r_{\rm min}^{2p})\right], \ee and
 \be L_{\nu}^{\rm
brem}=2.29\times10^{24}\alpha^{-2}c_{1}^{-2}m\dot{m}^{2}_{\rm out}
r^{-2p}_{\rm max}F(\theta_{e})T_{\rm e}^{-1}\times
\exp(-h\nu/\kappa T_{\rm e})\left[\frac{1}{2p}(r_{\rm
max}^{2p}-r_{\rm min}^{2p})\right], \ee
 where $F(\theta_{\rm e})$ is given in
M97 (Eq. 28).
 For the Comptonization of synchrotron photons,
the optical depth  $\tau_{\rm es}= \frac{1}{2}(\tau_{\rm es}^{\rm
max}+\tau_{\rm es}^{\rm min})$ is given by
 \be \tau_{\rm
es}=6.2\alpha^{-1}c_{1}^{-1}\dot{m}_{\rm out}r_{\rm
max}^{-p}\left[r_{\rm max}^{(2p-1)/2} + r_{\rm
min}^{(2p-1)/2}\right]. \ee The Compton spectrum and total Compton
power are given by \be L_{\nu}^{\rm comp}\simeq
L_{\nu_{i}}\left(\frac{\nu}{\nu_{i}}\right)^{-\alpha_{c}}, \ee and
 \be
P_{\rm comp}=\frac{\nu_{\rm p}L_{\nu_{\rm p}}^{\rm
sync}}{1-\alpha_{c}}\left[\left(\frac{6.2\times10^{10}T_{\rm
e}}{\nu_{\rm p}}\right)^{1-\alpha_{c}}-1\right]. \ee

For a given set of parameters: $m$, $\dot{m}_{\rm out}$, $\alpha$,
and $\beta$, the total heating of the electrons should be balanced
by the sum of the individual cool terms: $Q^{\rm ie}=P_{\rm
sync}+P_{\rm brem}+P_{\rm comp}$. The electron temperature $T_{\rm
e}$ can be calculated from the above energy equations, and then
the spectrum of the ADIOS flow  can be calculated, if all these
parameters are specified.

\section{Results}

\subsection{Relations between black hole masses and luminosities}

   The relations between the black hole masses and the continuum
   luminosities in different wavebands, for the sources in our
   sample, are plotted in Figs. 1 (radio), 2 (optical), and 3 (X-ray), respectively. As
   discussed in Sect. 3, the maximal continuum luminosities in different wavebands, for a
   pure ADAF model, can be calculated as function of black hole mass $M_{\rm
   bh}$, if the parameter $\alpha$ is specified. The maximal
   luminosities increase with the viscosity parameter $\alpha$, so
   the maximal luminosity
   of $L_{\lambda}^{\rm max}(\rm
  5\ GHz)$, $L_{\lambda}^{\rm max}(2\ {\rm keV})$, and $L_{\lambda}^{\rm max}(4400\rm\ \AA)$ for
   $\alpha=1$ are plotted in Figs. 1-3, respectively. In Fig. 1, we find that 15 sources in our sample have
   5 GHz
   luminosities less than the maximal radio luminosities
   predicted by the pure ADAF model (hereafter, these 15 sources are referred to as radio-weak
   AGNs). For these radio-weak AGNs, their radio emission is so weak that can be explained
   by the ADAF model. However, these sources, different from the conventionally defined radio-quiet
   AGNs, have their radio-loudness $\log_{10} R^{'}=2\sim5$ (their optical core emission
   is also very weak). In Figs. 2 and 3, we find that all these radio-weak
   sources have optical and X-ray luminosities less than that
   predicted by the pure ADAF model. It is found that some
   sources, with radio luminosities higher than the maximal
   luminosities predicted by the pure ADAF model, are situated below the
   maximal optical and X-ray luminosities predicted by the
   pure ADAF model in Figs. 2 and 3 (see the square sources in Figs. 1, 2, and 3).

\subsection{Distribution of Eddington ratio $\lambda L_{\lambda}^{B}/L_{\rm Edd}$}
    The distribution of the Eddington ratio, $\lambda L_{\lambda}^{B}/L_{\rm Edd}$, is plotted in Fig. 4. A bimodal
    distribution of $\lambda L_{\lambda}^{B}/L_{\rm Edd}$ is found for the whole sample. We calculated the statistical
    significance of this possible bimodality, using the KMM algorithm \citep{abz94}.
    The KMM test shows that the distribution of $\lambda L_{\lambda}^{B}/L_{\rm Edd}$
    is non-unimodal (with P-value = 0.003 $<$ 0.05) with null hypothesis. No selection
    effects
    were found that may lead to such a bimodal distribution of
     $\lambda L_{\lambda}^{B}/L_{\rm Edd}$ for this sample.
   \citet{mcf04}
   also found that a bimodal distribution of the dimensionless accretion rate $\dot{m}$, for
   a sample of radio galaxies and radio-loud quasars, which is similar to
   ours. \citet{bc04} proposed that
   sources with lower accretion rates can be explained if ADIOS flows are present.  All
   15 radio-weak sources in our sample have systematically lower
   $\lambda L_{\lambda}^{B}/L_{\rm Edd}$  ratios, which may be in either the ADAF or ADIOS states.

\subsection{Test ADAF model}

   There are 15 radio-weak sources in our
   sample, including four sources (NGC 3377, NGC 4342, NGC 4697 and
   NGC 5845) without measured optical nuclei, and two sources (NGC 4596 and NGC 5845) without measured X-ray emission.
   There are 10 radio-weak sources having all optical, radio, and X-ray emission data.
    The black hole masses of these 10 radio-weak sources are in the range
   $2.95\times10^{6}-2.0\times10^{9}M_{\odot}$. The ADAF spectrum  can be calculated, if the black hole
   mass $m$, the accretion rate $\dot{m}$, and the parameters
   $\alpha$, $\beta$ are specified (M97). The spectral energy
   distributions (SEDs) are similar for different black hole masses,
   if the accretion rate $\dot{m}$,  the parameters
   $\alpha$, and $\beta$ are fixed.  So, $M_{\rm
   bh}=10^{8}M_{\odot}$ is adopted in our spectral calculations for ADAFs, and this should not affect our main conclusions. The relation
between
    $\lambda L_{\lambda}(4400\rm\ \AA)$/$L_{\rm Edd}$ and $L_{\lambda}(2\ {\rm keV})/L_{\lambda}(\rm
  5\ GHz)$  is plotted in
  Fig. 5, for $\alpha=1$. The ADAF spectral calculations for
  different field strengths $\beta$, and accretion rates $\dot{m}$, are
  also plotted in Fig. 5.
     Compared with our spectral calculations, the magnetic field
     strength ($B\propto (1-\beta)^{1/2}P_{\rm tot}$) is required to be very
     small ($1-\beta \rightarrow 0$), for some radio-weak sources, if
     they are indeed in a pure ADAF state. Some sources (e.g., NGC
     4649) are even
     required to have
     $1-\beta<10^{-5}$.  The corresponding results for $\alpha=0.1$
     are presented in Fig. 6.

    The radio emission from the ADAF model depends sensitively on the
    magnetic field strength.
    The maximal luminosity of a pure ADAF, with
     such a weak magnetic field strength, can also
    calculated  as described
    in Sect. 3. It is found that 6 radio-weak sources (NGC 221,
     NGC 3608, NGC 4291, NGC 4459, NGC 4564 and NGC 4649) in our sample have
    radio luminosity higher than the maximal values predicted by
    the ADAF model, if they indeed have the very weak magnetic field
    strengths (very low values of $1-\beta$), as required by the ADAF spectral
    calculations in these sources (see Figs. 1 and
    5). For example, the spectral calculations require
    $1-\beta\simeq10^{-5}$ for NGC 4649, while its radio
    luminosity is higher than the maximal luminosity of a pure ADAF model, calculated for $1-\beta=10^{-5}$.

\subsection{Test ADIOS model}
 The accretion rate $\dot{m}$ is no longer constant for an ADIOS
 flow, as matter is carried away by the winds.
 For an ADIOS flow,
$\dot{m}=\dot{m}_{\rm out}(r/r_{\rm max})^{p}$ is assumed, and the
detailed spectral calculations for ADIOS flow are similar to those
for pure ADAF model, as described in Sect. 3. For $\alpha$=1, the
relations between $\lambda L_{\lambda}(4400\rm\ \AA)$/$L_{\rm
Edd}$ and $L_{\lambda}(2\ {\rm keV})/L_{\lambda}(\rm
  5\ GHz)$ for $p$=0.4 (mild winds), and $p$=0.99 (strong winds), are plotted in Figs. 7 and
  8, respectively. In Fig. 7, we find that extremely weak magnetic
field strengths ($1-\beta\rightarrow0$) are required to explain
the observed spectra for these 10 radio-weak sources, if they are
in a pure ADIOS state, which is similar to that from the pure ADAF
models. ADIOS flow with  strong winds ($p=0.99$) cannot explain
the SEDs of all of these radio-weak sources.

\subsection{ADAF+SD model}
  An ADAF may be present near the
  black hole, and it may transit to a cold standard disk (SD) beyond the
  transition radius $r_{\rm tr}$ \citep{e97}. \citet{c02}'s
  calculations indicated that the optical continuum emission from
  the inner ADAF region can be neglected, compared with that from
  the outer SD, if $r_{\rm tr}$ is around tens to several hundred Schwarzchild radii.
  The spectrum of the outer SD can be calculated, if the black hole mass $M_{\rm bh}$,
  accretion rate $\dot{m}$, and the transition radius $r_{\rm tr}$, are specified. It is found that the
  transition radii of these radio-weak sources are around $100\sim200\
  R_{\rm Schw}$, if $\dot{m}=10^{-3}$ is adopted (see Fig. 2).

\section{Discussion }
\subsection{Different accretion modes of LLAGNs}
    Most LLAGNs in our sample have strong radio emission,
    which is probably dominated by jet
    emission (see Fig. 1). However, the
    origin of the radio emission, from the radio-weak LLAGNs in
    this  sample (the triangles in the figures) is still unclear. The radio emission may be from either the jets
    (outflows) or the ADAFs (ADIOS flows), or from both, because the
    synchrotron emission of the hot electrons in the
    ADAFs (ADIOS flows) is important. For the radio-weak LLAGNs in our sample,
    their radio emission is fainter than the maximal radio
    luminosity predicted by the ADAF model. Thus, in principle,
    their radio emission can be explained as the synchrotron emission
    from the hot electrons in ADAFs, which is supported by the
    fact  that their X-ray and optical luminosities are also lower than
    the maximal values predicted by the ADAF models (see Figs. 2 and
    3).
  All these 10 radio-weak LLAGNs have relatively lower values of $\lambda
  L_{\lambda}^{B}/L_{\rm
     Edd}$ (see Fig. 4). The bimodal distribution of $\lambda L_{\lambda}^{B}/L_{\rm
     Edd}$ implies that these radio weak
     sources probably have ADAFs or ADIOS flows, while the sources
     in the population with higher values of $\lambda L_{\lambda}^{B}/L_{\rm
     Edd}$ only have standard thin disks.

\subsection{
 Can the radio emission from the radio-weak LLAGNs be explained by
radiatively inefficient accretion models?}

    The spectra of ADAFs can be calculated, if the black hole mass $m$, the
    dimensionless accretion rate $\dot{m}$, the magnetic field
    strength parameter $\beta$, and the viscosity parameter $\alpha$, are
    specified. We compare the observational data in radio, optical
    and X-ray wavebands with theoretical ADAF spectral calculations
   in Figs. 5 and 6, for different values of $\alpha$.
    It is found that extremely weak magnetic field strengths (e.g.,
    $1-\beta<10^{-5}$) are required  if these 10 radio-weak sources are in the pure ADAF
    state. However, \citet{plg02} found that the two-temperature ADAF
    structure is suppressed if no magnetic field is present in the
    ADAF. It is still unclear whether an ADAF can have such weak
    magnetic field strengths, because the magnetic field can be easily
    amplified through dynamo processes in accretion
    flows. This may imply that the radio emission of these
    radio-weak sources cannot be solely attributed to pure ADAFs.

  The ADAF radio emission depends sensitively on the magnetic field strength.
  If the magnetic fields of the flows in these radio-weak sources are so weak, as required
  by the ADAF spectral calculations, we find that 6 of the 10 radio-weak sources have higher radio
  luminosities than the maximal luminosities for ADAFs (see Fig. 1). This indicates that
  the ADAFs are unable to produce the observed radio emission, at least in these six sources, even if
  the field strength is extremely weak. Thus, jet radio emission may be important in
  these sources, which is consistent with the two very weak radio lobes
    detected in NGC 4649 \citep{sw86}.

     Similar analyses have been done for the ADIOS cases.
     The results of the ADIOS models with mild winds are similar to the ADAF cases (Fig. 7). Very
     weak magnetic fields are required in the ADIOS flows. The
     ADIOS flows with strong winds, for example, $p=0.99$, cannot
     reproduce the observed radio emission of all these radio-weak
     sources.

     Convection dominated accretion flows (CDAFs) may be a
     possible
     alternative for ADAFs or ADIOS flows \citep{ni00, spb99}.
     The accretion rates of CDAFs are much smaller than those in
     non-convecting ADAFs, so the CDAFs are relatively very faint \citep{bnq01}. This implies
     that most radio-weak LLAGNs in this sample may be too bright
     to be powered by pure CDAFs, though the quantitatively
     calculations are still unavailable. Even if CDAFs are present
     in these sources, we expect that the jet emission is still required to
     explain their observed radio emission, and our main conclusions on
     the origin of their radio emission should remain unchanged.

     In our analyses, we have assumed that all the emission
     in different wavebands (radio, optical, and X-ray) is from the nuclei of the
     sources,
     without being absorbed. We note that several sources in our sample are
     type 2 AGNs, and that putative
     tori may be present to obscure nuclear emission. The $B$-band
     luminosities of these sources are estimated by using the
     relation $L_{\rm H\beta}- M^{'}_{B}$ \citep{hp01}. For type 2
     AGNs, only narrow-line emission is observed, as broad-line
     emission may be obscured by the putative tori. It is found that
     the mean ratio of the broad to narrow
      components $\rm H\beta(\rm broad)/\rm H\beta(\rm narrow)$ is $\sim 7.1$, for a sample of 119 low-redshift
     Seyfert 1 type AGNs \citep{x03}.  \citet{os81} found that
     the ratio $\rm H\beta(\rm broad)/\rm H\beta(\rm narrow)$ is around 2-4.5 for five Seyfert
     1 galaxies.  \citet{skk94} also found that this ratio is
     around 1.6-7 for three low-luminosity Sefert 1 galaxies. Even
     if this is the case for radio-weak sources in our sample, the
     $B$-band luminosities of any type 2 LLAGNs may be
     underestimated by no more than one order of magnitude, and most
     of the radio-weak LLAGNs in our sample still cannot be interpreted with reasonable magnetic field
     strengths (Figs. 5-8). If the putative tori are indeed present
     to obscure the nuclear emission, the 2 keV X-ray emission
     may be partly absorbed by the tori, if not all absorbed.
     In this case, the intrinsic X-ray luminosities of type 2
     LLAGNs should higher than the observed values. This
     requires the magnetic field strengths of the accretion flows to be even
     weaker than in our present calculations (see Figs. 5-8).
     Thus our main conclusion on the origin of the radio emission will
     remain unchanged, even if the nuclear emission of some radio-weak
     LLAGNs may be obscured by the putative tori.

\subsection{Other possible models for radio-weak LLAGNs}

     It should be cautioned that our present analyses are performed
     by assuming all the emission in the different wavebands to be
     from ADAFs or ADIOS flows. If the ADAF (ADIOS flow) is
     truncated to a standard thin disk at a certain radius, the
     observed optical emission may be dominated by the
     outer standard disk. For such ADAF+SD (or ADIOS+SD) systems, it is found that the transition radii
     of these radio-weak LLAGNs should be around $100 \sim 200\ R_{\rm
     Schw}$,
     if $\dot{m}=10^{-3}$ is adopted (Fig. 2). If this is the case, the optical
     emission from the ADAF (or ADIOS flow) should be less than
     the observed values. This requires the points in Figs.5-8 to
     be shifted towards the left, i.e., much weaker field strengths
     are
     required in the spectral calculations and our conclusions
     will thus be strengthened.

      Besides the ADAF or
      ADIOS flow, the X-ray emission may also be from the jet or
      the corona above the outer standard thin disk surrounding the ADAF
      or ADIOS flow. In this case, only a fraction of the
      observed X-ray emission is from the ADAF or ADIOS flow. The
      locations of the sources would be shifted downwards, and they could be explained
       by the ADAF/ADIOS flow with relatively
      stronger magnetic field strength (see Fig. 5).
     Comparisons of the observed optical luminosities with the
     theoretical spectral calculations for ADAF+SD systems require
     the transition radii of these radio-weak LLAGNs to be larger
     than 100 $R_{\rm Schw}$, if $\dot{m}=10^{-3}$ is adopted. For
     a higher adopted $\dot{m}$, the transition radii should be
     larger. \citet{wcz97} found that
     the corona cannot survive if the accretion rate is less than a critical value
      at a given radius (Figs. 3a, b in their paper). Their
      calculations showed that, the corona should be absent if the
      transition radii of these radio-weak LLAGNs are larger than 100 $R_{\rm
      Schw}$, with such a low accretion rate. Thus, the X-ray
      emission cannot be attributed to the corona above the outer
      thin disk for these radio-weak LLAGNs. The radio flux
      densities of these radio-weak sources are around
      sub-mJy, except for NGC 4649 and Milky way, which are beyond the observational capability
      of VLBI. We suggest that future high-sensitivity Square Kilometer Array (SKA) observations
     of these LLAGNs  should help to verify the origins of their radio emission.

    \acknowledgments We thank  K. A. Ashman, and  S. E. Zepf,
 for providing us the  KMM code, and L. C. Ho for helpful
discussion.
    The authors are grateful to the referee for constructive suggestions on our
    paper. We also thank A. B. Fletcher (SHAO) for help with
    proofreading and discussing this paper.
    This work is supported by the National Science
     Fund for Distinguished Young Scholars (grant 10325314), NSFC
     (grants 10173016; 10333020), and the NKBRSF (grant G1999075403).
     This research has made use of the NASA/IPAC Extragalactic Database
    (NED), which is operated by the Jet Propulsion Laboratory,
     California Institute of Technology, under contract with the
     National Aeronautic and Space Administration.

\clearpage
\begin{deluxetable}{lcccrcc}
\tabletypesize{\scriptsize}
 \tablecaption{Sample Data } \tablewidth{0pt} \tablehead{
\colhead{Source Name} & \colhead{$\log_{10}\ m ^{\rm a}$}&
\colhead{$\log_{10} L_{\lambda}{(4400\  \rm \AA)}$ $^{\rm b}$}  &
\colhead{$\log_{10} L_{\lambda}(\rm 5\ GHz)$ $^{\rm c}$} &
\colhead{ Distance $^{\rm d}$} & \colhead{$\log_{10}
L_{\lambda}(\rm 2\ keV)$ $^{\rm e}$} & \colhead{References} }

\startdata
$\rm Circles^{f}$                    \\
                           \\
3C 120 (Mrk 1506)   & 7.36   & 40.20 & 24.90  & 137.8 & 40.40  & 1 \\
3C 390.3 (VII w 838)& 8.53   & 40.33 & 25.62  & 241.2 & 40.49  & 2 \\
Ark 120 (Mrk 1095)  & 8.26   & 40.93 & 21.80  & 134.6 & 40.45  & 1 \\
Arp 102B            & 8.34   & 38.92 & 22.87  & 99.7  & 39.83  & 3 \\
Circinus            & 6.11   & 39.56 & 21.04  & 4.0   & 38.49  & 4 \\
Fairall 9           & 7.90   & 41.13 & 22.96  & 199.8 & 40.34  & 1 \\
IC4329A             & 6.69   & 39.58 & 22.25  & 65.5  & 39.94  & 1 \\
Mrk 79 (UGC 3973)   & 7.71   & 39.85 & 21.62  & 91.3  & 39.81  & 1  \\
Mrk 110             & 6.74   & 39.64 & 21.76  & 147.7 & 40.25  & 5 \\
Mrk 279 (UGC 8823)  & 7.62   & 39.84 & 22.09  & 126.6 & 40.21  & 2 \\
Mrk 335             & 6.79   & 40.23 & 21.60  & 106.6 & 39.78  & 1 \\
Mrk 509             & 7.76   & 40.87 & 22.11  & 143.8 & 40.51  & 1 \\
Mrk 590 (NGC 863)   & 7.25   & 38.85 & 21.91  & 109.2 & 40.05  & 1 \\
Mrk 817 (UGC 9412)  & 7.64   & 39.96 & 22.05  & 131.0 & 39.22  & 6 \\
NGC 1068 (M77)      & 7.20   & 38.47 & 22.43  & 14.4  & 37.50  & 1 \\
NGC 3227            & 7.59   & 37.84 & 21.20  & 20.6  & 38.44  & 1 \\
NGC 3516            & 7.36   & 39.07 & 21.34  & 38.9  & 39.69  & 2 \\
NGC 3783            & 6.97   & 39.48 & 21.36  & 38.5  & 39.20  & 1 \\
NGC 4051            & 6.11   & 37.63 & 20.54  & 17.0  & 38.34  & 1 \\
NGC 4151            & 7.18   & 39.30 & 21.84  & 20.3  & 39.28  & 1 \\
NGC 4593            & 6.90   & 39.00 & 20.48  & 39.5  & 39.21  & 1 \\
NGC 4945            & 6.04   & 37.64 & 21.59  & 4.2   & 38.74  & 4 \\
NGC 5548            & 8.09   & 39.71 & 21.86  & 70.2  & 39.68  & 1 \\
NGC 7469            & 6.81   & 39.82 & 22.47  & 66.6  & 39.64  & 1 \\
                                                            \\
$\rm Squares^{g}$                           \\
                                                      \\
IC 342              & 5.70   & 33.48 & 17.95  & 1.8   & 34.84  & 7 \\
IC1459              & 8.57   & 36.66 & 23.02  & 29.2  & 37.15  & 2 \\
NGC 205 (M110)      & 4.97   & 31.04 & 17.12  & 0.74  & 32.19  & 8 \\
NGC 598 (M33)       & 3.18   & 31.29 & 16.70  & 0.87  & 34.97  & 7  \\
NGC 821             & 7.70   & 35.05 & 19.54  & 24.1  & 37.75  & 9 \\
NGC 1023            & 7.59   & 34.62 &   ...  & 11.4  &  ...   & ...\\
NGC 2778            & 7.30   & ...   & 19.58  & 22.9  & 36.95  & 9  \\
NGC 2787            & 7.59   & 35.19 & 19.78  & 7.5   & 34.89  & 2 \\
NGC 3031 (M81)      & 7.80   & 36.12 & 20.23  & 3.9   & 36.73  & 10 \\
NGC 3245            & 8.32   & 36.36 & 20.24  & 20.9  &  ...   & ...\\
NGC 3384            & 7.25   & 34.60 & 18.91  & 11.6  & 36.00  & 11 \\
NGC 3998            & 8.75   & 37.06 & 21.29  & 14.1  & 38.15  & 2 \\
NGC 4203            & 7.08   & 36.11 & 20.43  & 14.1  & 36.29  & 12 \\
NGC 4258 (M106)     & 7.61   & 35.11 & 19.35  & 7.3   & 35.06  & 12 \\
NGC 4261            & 8.72   & 35.99 & 22.58  & 31.6  & 37.66  & 2 \\
NGC 4374 (M84)      & 9.20   & 35.67 & 22.15  & 18.4  & 36.14  & 7 \\
NGC 4395            & 5.04   & 34.87 & 18.29  & 3.6   & 34.04  & 12 \\
NGC 4473            & 8.00   & 34.75 & 19.77  & 15.7  & 36.04  & 13 \\
NGC 4486 (M87)      & 9.53   & 36.23 & 23.09  & 16.1  & 37.42  & 12 \\
NGC 4594 (M104)     & 9.04   & 36.00 & 21.15  & 9.8   & 37.19  & 2  \\
NGC 6251            & 8.73   & 36.24 & 24.22  & 94.8  & 38.64  & 2  \\
NGC 7052            & 8.56   & 36.75 & 22.64  & 63.6  &  ...   & ...\\
NGC 7457            & 6.53   & 33.91 &   ...  & 13.2  & 36.50  & 9  \\

                                                      \\
$\rm Triangles^{h}$           \\
                            \\
MilkyWay            & 6.47   & 29.07 & 15.79  & 0.008  & 28.61  & 14 \\
NGC 221 (M32)       & 6.59   & 32.73 & 16.89  & 0.81   & 34.14  & 7  \\
NGC 224 (M31)       & 7.52   & 36.02 & 15.40  & 0.76   & 35.81  & 15 \\
NGC 3115            & 8.96   & 34.92 & 18.57  & 9.7    & 35.44  & 13 \\
NGC 3377            & 8.00   &  ...  & 18.88  & 11.2   & 35.53  & 11 \\
NGC 3379 (M105)     & 8.00   & 34.98 & 19.03  & 10.6   & 34.92  & 12 \\
NGC 3608            & 8.04   & 35.08 & 19.50  & 22.9   & 36.29  & 11 \\
NGC 4291            & 8.17   & 35.09 & 19.43  & 26.2   & 37.01  & 11 \\
NGC 4342            & 8.53   &  ...  & 19.23  & 16.8   & 35.65  & 13 \\
NGC 4459            & 7.81   & 35.58 & 19.39  & 16.1   & 35.95  & 13 \\
NGC 4564            & 7.75   & 34.44 & 19.13  & 15.0   & 35.90  & 11 \\
NGC 4596            & 7.76   & 34.89 & 19.23  & 16.8   &  ...   & ...\\
NGC 4649 (M60)      & 9.30   & 34.44 & 20.78  & 16.8   & 37.22  & 11 \\
NGC 4697            & 8.08   &  ...  & 19.01  & 11.7   & 36.18  & 11 \\
NGC 5845            & 8.51   &  ...  & 19.60  & 25.9   &  ...   & ... \\

\enddata
\tablenotetext{a}{ $m=M_{\rm bh}/M_{\odot}$} \tablenotetext{b}{ in
unit of $\rm erg\ s^{-1}\ \AA^{-1}$} \tablenotetext{c}{ in unit of
$\rm W\ Hz^{-1}$}  \tablenotetext{d}{in unit of Mpc}
\tablenotetext{e}{ in unit of $\rm erg\ s^{-1}\ eV^{-1}$}
\tablenotetext{f}{ The circles represent sources with
$L_{\lambda}(B) > L_{\lambda}^{\rm max}(B)$ predicted by the ADAF
models (see Fig. 2). }\tablenotetext{g}{ The squares represent
sources with optical luminosity $L_{\lambda}(B) < L_{\lambda}^{\rm
max}(B)$ but with $L_{\lambda}(5\rm\ GHz)>$ $ L_{\lambda}^{\rm
max}(\rm 5\ GHz)$ predicted by the ADAF models (see Figs. 1 and
2). }\tablenotetext{h}{ The triangles represent the sources with
radio luminosity $L_{\lambda}(5\rm\ GHz)$$ < L_{\lambda}^{\rm
max}(\rm 5\ GHz)$ predicted by the ADAF models (see Fig. 1). }

 \tablerefs{(1)\citet{tp89}; (2)\citet{mhd03};
(3)\citet{ehc03}; (4)\citet{mf00}; (5)\citet{g00}; (6)\citet{p01};
(7)\citet{cm99}; (8)\citet{zm01} ; (9)\citet{b99};
(10)\citet{i96};
 (11)\citet{fkt92}; (12)\citet{h01}; (13)\citet{wga00}; (14)\citet{b03}; (15)\cite{t99}.  }
\end{deluxetable}

\clearpage
\begin{figure}
\plotone {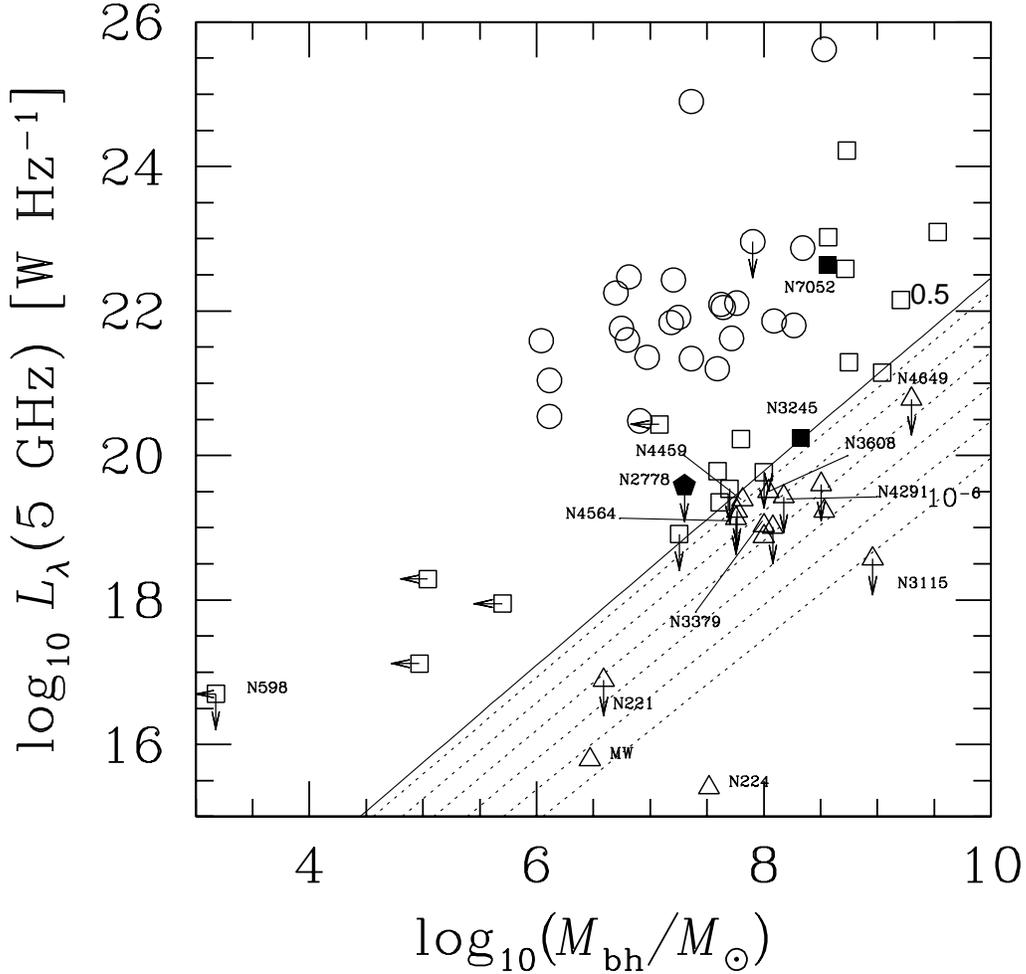}

 \caption{The relation between the black hole mass $M_{\rm bh}$
 and radio core luminosity $L_{\lambda}$ at 5 GHz. The triangles
represent the sources with radio luminosity $L_{\lambda}(5\ \rm
GHz)$ $< L_{\lambda}^{\rm max}(5\ \rm GHz)$ predicted by the ADAF
models. The solid line represents the maximal radio luminosity of
the ADAF model at 5 GHz, as a function of black hole mass, for
$\beta$=0.5. The dotted lines represent the maximal radio
luminosity of the ADAF model for
$1-\beta=10^{-1},10^{-2},10^{-3},10^{-4},10^{-5}$, and $10^{-6}$
(from top to bottom), respectively. A viscosity of $\alpha=1$ is
adopted in the calculations. The squares represent the sources
with optical luminosity $L_{\lambda}(B) < L_{\lambda}^{\rm
max}(B)$ predicted by the ADAF models, but also with
$L_{\lambda}(5\rm\ GHz)$ $ > L_{\lambda}^{\rm max}(\rm 5\ GHz)$
(see Fig. 2). The circles represent the sources with
$L_{\lambda}(B) > L_{\lambda}^{\rm max}(B)$ predicted by the ADAF
models. The two black square sources NGC 3245 and NGC 7052 have
not been detected in the X-ray waveband, while the optical nucleus
of the black pentagon source NGC 2778 has not been detected.
 \label{fig1}}

\end{figure}

\begin{figure}
\plotone {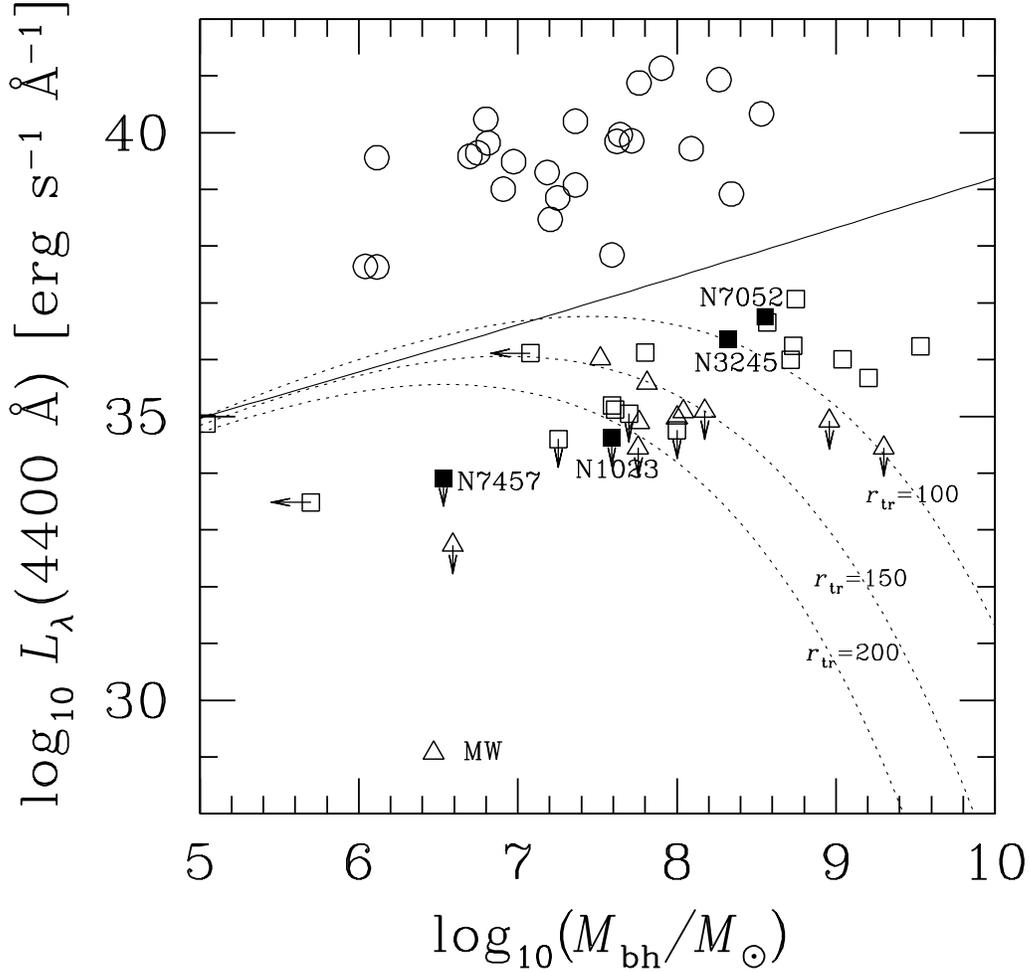}

\caption{The relation between the black hole mass $M_{\rm bh}$
 and optical core luminosity $L_{\lambda}$ at $B$ band. The solid line
 represents the maximal optical luminosity of the ADAF model at $B$ band as a function
of black hole mass. The dotted lines represent ADAF+SD models with
the different transition radii $r_{\rm tr}$, for
$\dot{m}=10^{-3}$. No radio emission data are available for NGC
1023 and NGC 7457, while no X-ray emission data are available for
NGC 3245 and NGC 7052. \label{fig2}}

\end{figure}

\begin{figure}
\plotone {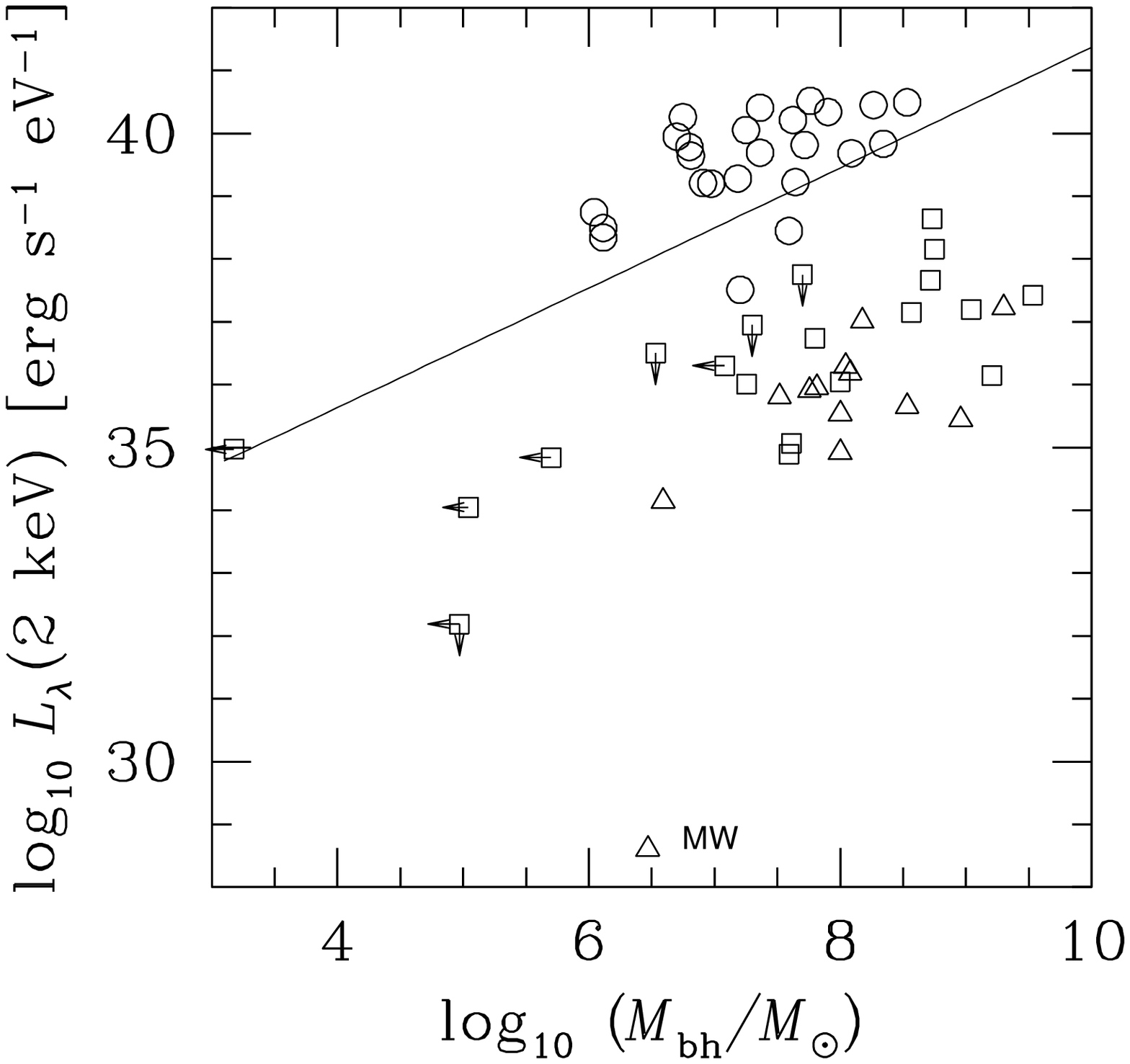}

 \caption{The relation between the black hole mass $M_{\rm bh}$
 and X-ray luminosity $L_{\lambda}$ at 2 keV. The solid line
 represents the maximal X-ray luminosity of ADAF at 2 keV, as a function
 of black hole mass. \label{fig3}}

\end{figure}

\begin{figure}
\plotone {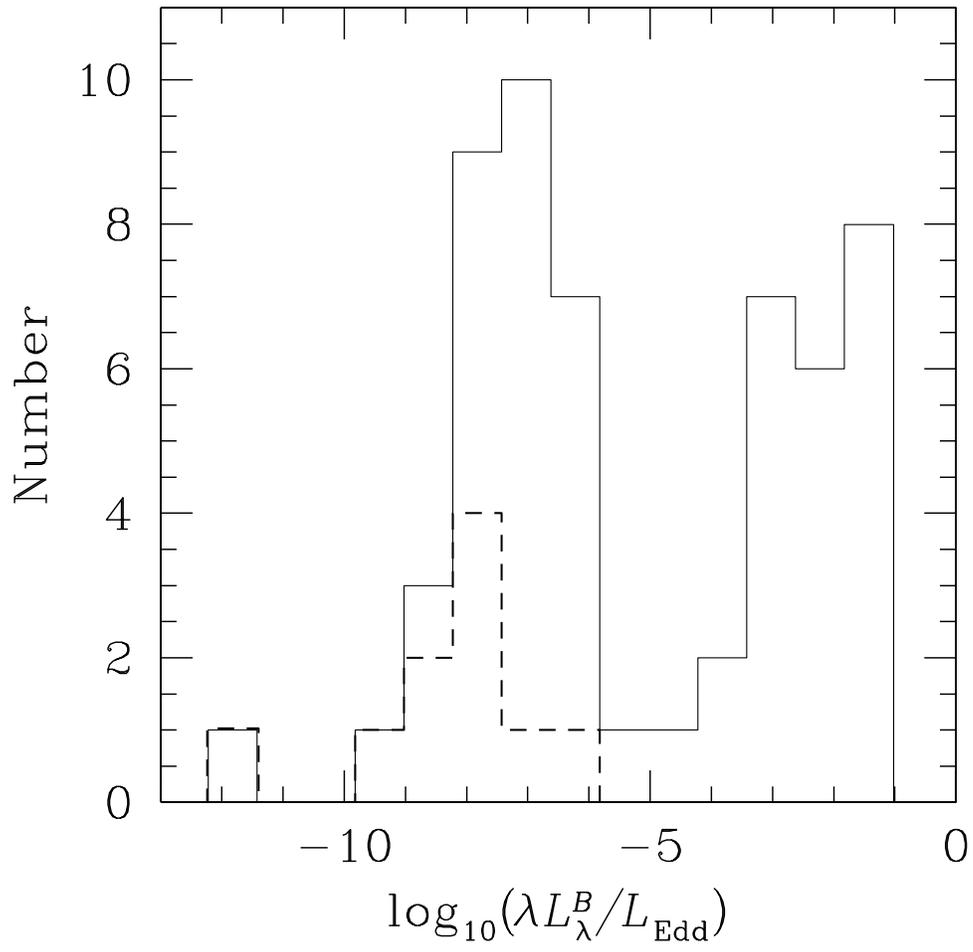}

\caption{The distribution of $\log_{10}(\lambda
L_{\lambda}^{B}/L_{\rm Edd})$ for the sources in our sample (solid
line: all sources; dashed line: radio-weak sources). \label{fig4}}
\end{figure}

\begin{figure}
\plotone {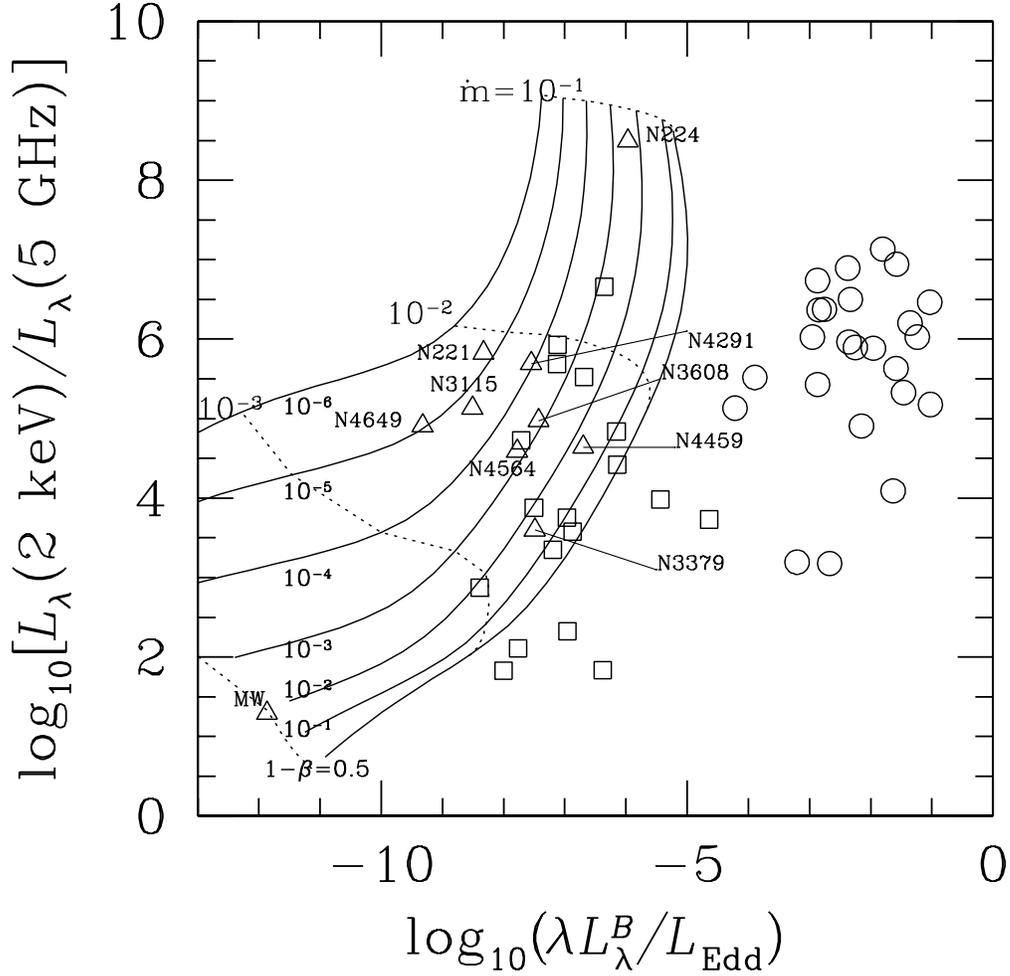}

\caption{The relation between $\lambda L_{\lambda}^{B}/L_{\rm
Edd}$ and $L_{\lambda}(2\rm\  keV)$/$L_{\lambda}(5\rm\  GHz)$. The
solid lines represent results from spectral calculations for ADAFs
with different values of $1-\beta$ (from 0.5 to $10^{-6}$). The
dotted lines represent results from spectral calculations with
different values of $\dot{m}$ (from $10^{-4}$ to $10^{-1}$). The
calculations are carried out for $M_{\rm bh}=10^{8}M_{\odot}$ and
$\alpha=1$. \label{fig5}}

\end{figure}

\begin{figure}
\plotone {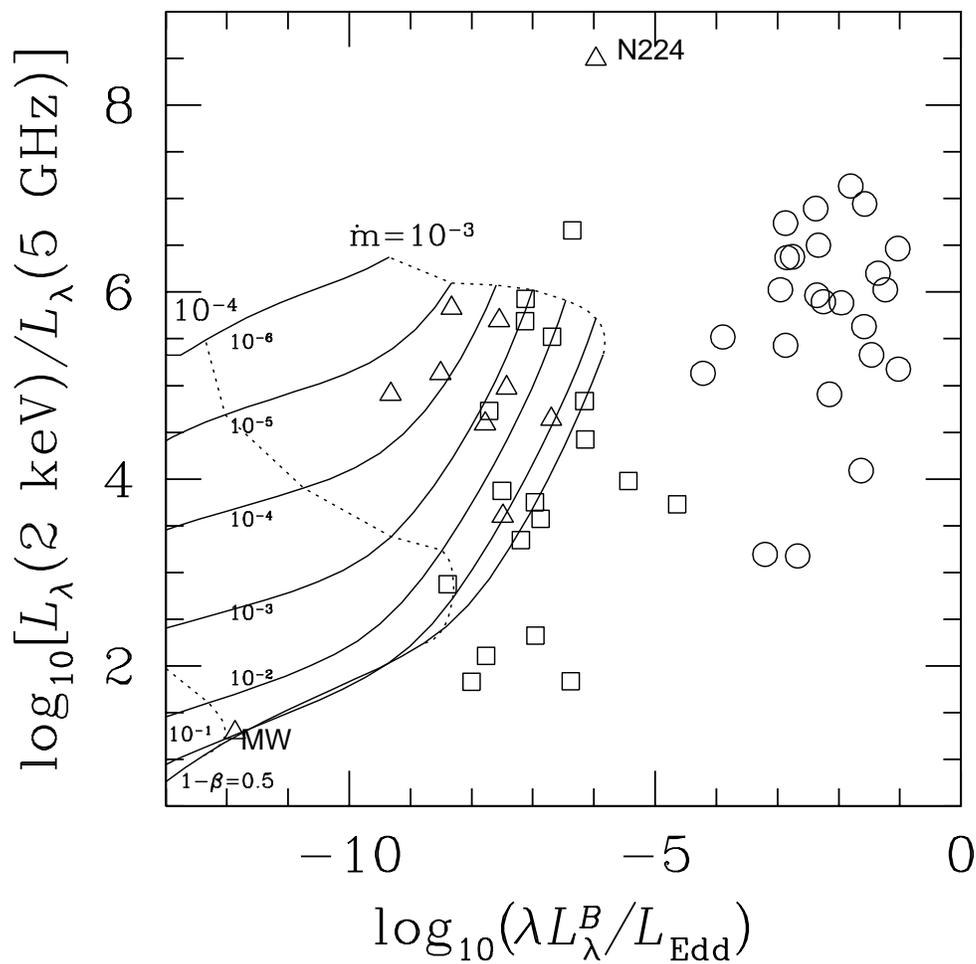}

\caption{Same as in Fig. 5, but $\alpha=0.1$ is adopted. The
intersection of the solid lines (for $1-\beta$ = 0.1 and 0.5) in
this Figure arises because we only calculate the ratio
$L_{\lambda}(2\rm\ keV)$/$L_{\lambda}(5\ \rm GHz)$. \label{fig6}}
\end{figure}

\begin{figure}
\plotone {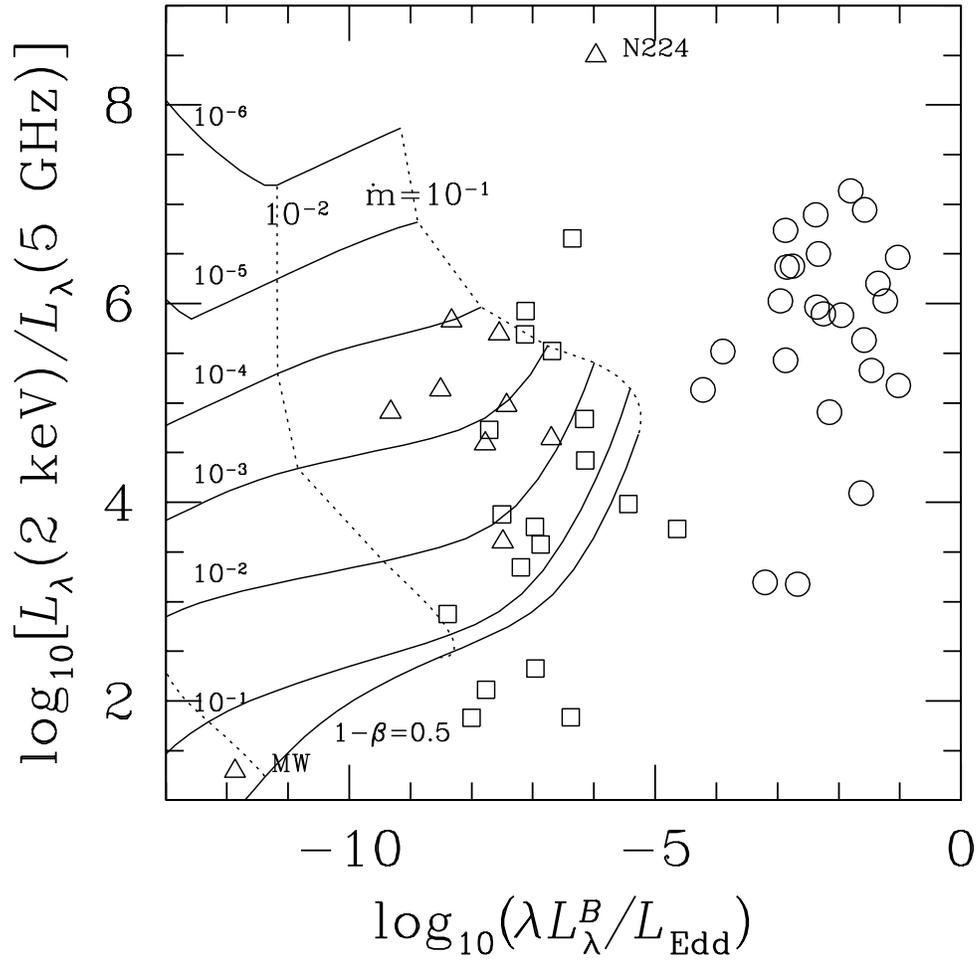}

\caption{Same as in Fig. 5, but for ADIOS models with mild winds
($p$ = 0.4). \label{fig7}}
\end{figure}

\begin{figure}
\plotone {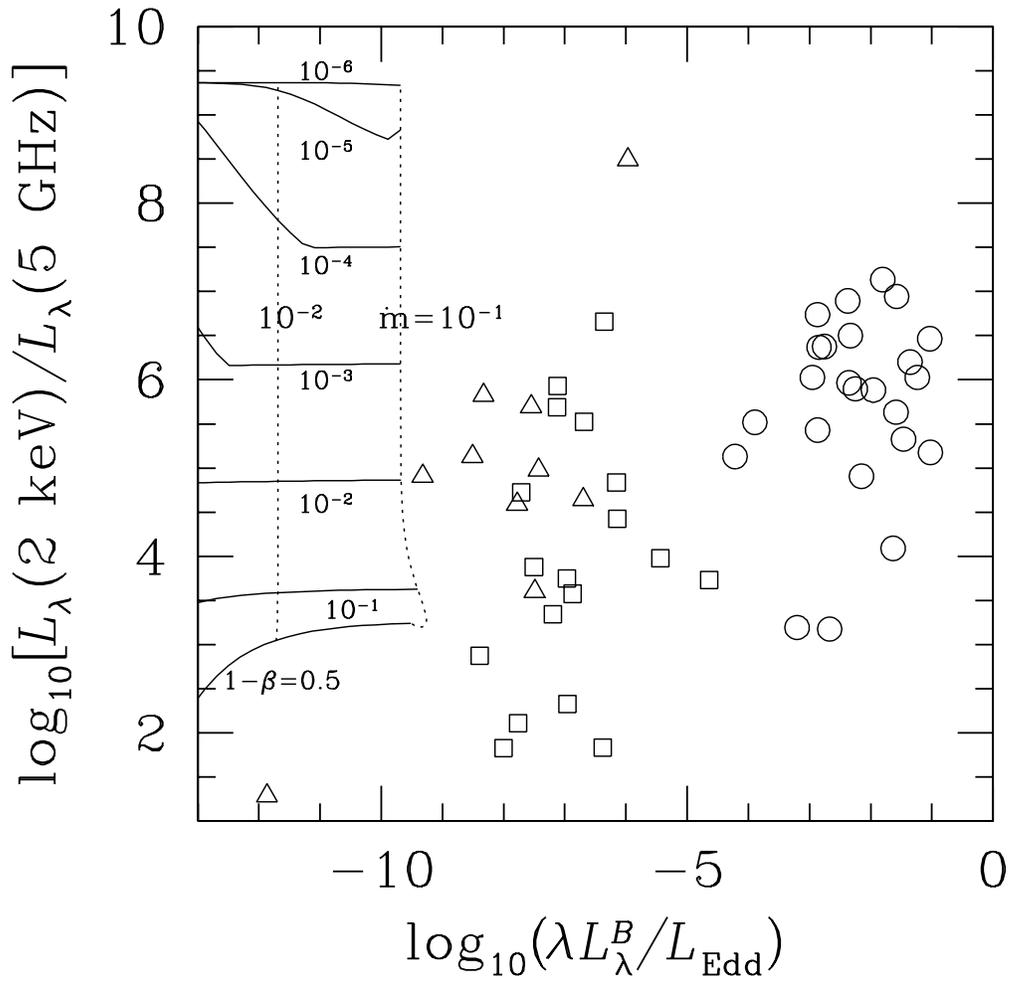}

\caption{Same as in Fig. 5, but for strong winds ($p$ = 0.99).
\label{fig8}}
\end{figure}

\end{document}